\documentclass[12pt]{article}
\usepackage{amsmath,amssymb}
\begin{document}

\begin{center}
{\Large\bf Lorentz-Invariant Non-Commutative QED}\\[0.8cm]
Katsusada Morita\\[0.2cm]
{\it  Department of Physics, 
  Nagoya University, Nagoya 464-8602, Japan}\\[0.3cm]
(\today)
\end{center}
\vspace*{0.6cm}

\begin{abstract}
{\normalsize
Lorentz-invariant non-commutative QED (NCQED) is constructed such that
it should be a part of Lorentz-invariant non-commutative 
standard model (NCSM), a subject to be treated in later publications.
Our NCSM is based on Connes' observation that the total fermion field in the
standard model may be regarded as a bi-module over a flavor-color
algebra. In this paper, it is shown that there exist
two massless gauge fields in NCQED which are interchanged
by $C'$ transformation. Since $C'$ is reduced to the conventional
charge conjugation $C$ in the commutative limit,
the two gauge fields become identical to the photon field
in the same limit, which couples to only four spinors with charges $\pm 2,\pm 1.$
Following Carlson-Carone-Zobin, our NCQED
respects Lorentz invariance employing Doplicher-Fredenhagen-Roberts
algebra instead of the usual algebra with constant $\theta^{\mu\nu}$.
In the new version $\theta^{\mu\nu}$ becomes an integration variable.
We show using a simple NC scalar model 
that the $\theta$ integration gives an {\it invariant} 
damping factor instead of the oscillating one
to the nonplanar self-energy diagram in the one-loop approximation.
Seiberg-Witten map shows that the $\theta$ expansion of NCQED
generates exotic but well-motivated 
derivative interactions beyond QED 
with allowed charges being only $0, \pm 1, \pm 2$.}
\end{abstract}
\vspace*{5cm}
\hspace*{1cm}
e-mail: morita@eken.phys.nagoya-u.ac.jp

\newpage

\section{Introduction}

This is the first in a series of papers
which are devoted to discuss the standard model on
non-commutative space-time.
The non-commutative
standard model (NCSM) was already taken up by
several authors$^{[1],[2],[3]}$. Our approach is different
from theirs in that
their model building is affected in no way by 
Connes' reformulation$^{[4],[5]}$ of the standard model, 
while we are motivated by the assumption$^{[6]}$ 
that {\it an (associative) algebra underlies the gauge symmetry},
and their non-commutative space-time with constant $\theta^{\mu\nu}$
breaks Lorentz symmetry, while 
we maintain Lorentz invariance 
following Carlson, Carone and Zobin$^{[7]}$.
\\
\indent
Let us first recall the main reorganization
introduced by Connes' intrusion.
Consider $U(1)$ gauge theory.  Connes
realized it by the algebra ${\cal A}=C^\infty(M_4)\otimes{\bf C}$ 
whose unitaries constitute $U(1)$.
From the linearity of the algebra representation $\rho(a+b)=\rho(a)+\rho(b)$
the allowed representation of the algebra 
is restricted to be of the form
$\rho(b)={\rm diag}\;(b,\cdots,b^{*},\cdots)$
for $b=b(x)\in{\cal A}$,
meaning the Abelian charge to be $\pm 1$ in Connes' realization
of $U(1)$ gauge theory$^{[8]}$, since the gauge group
is given by
the unitary group of the algebra, ${\cal U}(C^\infty(M_4)\otimes {\bf C})
)=$Map$(M_4,U(1))$.
\\
\indent
Next consider leptons. Since they have flavor and
the Abelian charges $Y(e_R)=-2, Y(l_L)=-1, Y(\nu_R)=0$ (assuming the right-handed neutrino),
we represent them as a bi-module.
In general, in an $A$-$B$ bi-module $M$ the two {\it commuting}
operations are defined,
$$
a(bx)=b(ax),\;\;\;a\in A,\;\;\;b\in B,\;\;\;x\in M.
$$
It is common to write one operation as a right action,
$$
a(xb)=(ax)b,\;\;\;a\in A,\;\;\;b\in B,\;\;\;x\in M.
$$
Similarly, we write the standard gauge transformation 
in the lepton sector in the form,
\begin{eqnarray}
^g\psi&=&(g\psi)u^{*}=g(\psi u^{*})\equiv g\psi u^{*},\nonumber\\[2mm]
g&=&\left(
                               \begin{array}{cc}
                               g_L&0\\
                               0&g_R\\
                               \end{array}
                               \right)\otimes 1_{N_g},\;\;\;g_R=\left(
                               \begin{array}{cc}
                               u&0\\
                               0&u^{*}\\
                               \end{array}
                               \right),
\end{eqnarray}
where $g_L\in SU(2)_L$, $u=e^{i\alpha}$
and $\psi=\left(
             \begin{array}{l}
             \psi_L\\
             \psi_R\\
             \end{array}
             \right)$ with $\psi_L=\left(
    \begin{array}{l}
    \nu\\
    e\\
    \end{array}
    \right)_L$ and $\psi_R=\left(
                             \begin{array}{l}
                             \nu_R\\
                             e_R\\
                             \end{array}
                             \right)$
in $N_g$ generations.
We distinguish between the left-handed doublets and the right-handed singlets
by the subscripts, $L$ and $R$, outside and inside the vector notation,
{\small$\left(
\begin{array}{l}
\!\!\!\\
\end{array}
\right)$}, 
respectively.
The form $u=e^{i\alpha}$ is due to the normalization
$Y(\phi)=1$. (See below.) Since $g$ and $u^*$ commute,
the leptonic $\psi$ is a bi-module over a flavor algebra,
$C^\infty(M_4)\otimes ({\bf H}\oplus{\bf C})$, ${\bf H}$ being real quaternions.
The unitary
group of the algebra, ${\cal U}(C^\infty(M_4)\otimes ({\bf H}\oplus{\bf C})
)=$Map$(M_4,SU(2)\times U(1))$ is the flavor group.
\\
\indent
Let us now introduce quarks into the scheme.
The total fermion field
\footnote{
This name was suggested by I. S. Sogami$^{[9]}$to
represent chiral leptons and quarks in a unified way.}
\begin{eqnarray}
\psi&=&\left(
    \begin{array}{cccc}
    l_L&q_L^r&q^b_L&q^g_L\\
    l_R&q_R^r&q^b_R&q^g_R\\
    \end{array}
    \right),
\end{eqnarray}
receives the standard gauge transformation,
\begin{eqnarray}
\psi\to ^g\!\psi=(g\psi)G=g(\psi G)\equiv g\psi G,
\end{eqnarray}
where the left action by $g$ is the same as in (1), while
the color gauge transformation is written as the right action by
\begin{eqnarray}
G&=&\left(
    \begin{array}{cc}
    u^*&0\\
    0&v^T\\
    \end{array}
    \right).
\end{eqnarray}
Here $v$ belongs to the unitary group of the color
algebra $C^\infty(M_4)\otimes M_3({\bf C})$.
Since no algebra exists whose unitary group is color $SU(3)$,
we need Connes' {\it unimodularity condition}
to reproduce the correct hypercharge of quarks,
\begin{eqnarray}
{\rm det}\;G=1.
\end{eqnarray}
Putting $v=e^{i\beta}v'$ with det$\;v'=1$
the unimodularity condition (5)
implies
\begin{eqnarray*}
-\alpha+3\beta=0.
\end{eqnarray*}
This correctly reproduces the fractional
hypercharge of quarks, $Y(q_L)=0+1/3=1/3, Y(u_R)=1+1/3=4/3, Y(d_R)=-1+1/3=-2/3$.
(3) defines the total fermion field as a bi-module
\footnote{
One can easily reassemble $\psi$ so as to move the right action
to the left with commutativity from semi-simple group structure becoming apparent.}.
On the other hand, the gauge transformation
of Higgs $h=\left(
  \begin{array}{cc}
  \phi_0^{*}&\phi_+\\
  -\phi_-&\phi_0\\
  \end{array}
 \right)$ looks like
\footnote{
The hypercharge of Higgs $\phi$ is normalized to be $+1$,
leading to the choice $u=e^{i\alpha}$.}
\begin{eqnarray}
h&\to& ^gh=(g_Lh)g_R^{\dag}=g_L(hg_R^{\dag})\equiv g_Lhg_R^{\dag}.
\end{eqnarray}
Consequently, matrix-valued Higgs is also regarded as a (single) bi-module.
The spontaneous breakdown of symmetry is triggered by
the finite vacuum expectation value $\langle h\rangle=(v/\sqrt{2})1_2$ 
so that it is given by
\begin{equation}
g_L\to g_R.
\end{equation}
\indent
In a series of papers we interpret the bi-module structure
of the total fermion field in the framework of
non-commutative gauge theory
(NCGT), defining the total fermion field as a non-commutative (NC) bi-module
so that the round brackets in (1) and (3) mean only the associativity
\footnote
{For instance, we can no longer write (1) as if both actions
operate on the spinor only from the left. In particular, this means 
that $e_R$ and $\nu_R$ remain acted from both sides, 
$e_R\to ^g\!\!e_R=u^{*}e_Ru^{*}$ and $\nu_R\to ^g\!\!\nu_R=u\nu_Ru^{*}$,
which would read $^ge_R=u^{*}u^{*}e_R=(u^*)^2e_R$ 
and $^g\nu_R=uu^{*}\nu_R=\nu_R$ in the standard notation, respectively.
If we start with this notational convention,
the left-handed doublet and the right-handed singlet
behave `differently' on non-commutative space-time as assumed in Ref.1).
(In Refs. 1) and 2) $\nu_R$ is not considered.)
}.
We are motivated to study NCSM
for two reasons.
Firstly, it was shown in Ref. 10)
that,
in non-commutative QED
(NCQED), Abelian charge is restricted to be $\pm 1$ and 0.
This is similar to the restriction in Connes' realization of $U(1)$ gauge theory.
If we consider NCQED as only a part of
a larger theory, NCSM, it is necessary to
incorporate the value $\pm 2$ in NCQED
to account for the Abelian charge $Y(e_R)=-2$.
This is accomplished by considering NC bi-module 
in exactly the same way as we explained $Y(e_R)=-2$
by considering the bi-module. We are thus led to the second
motivation, namely, if the field quantities are defined on
non-commutative space-time, the left and right actions are distinguished,
interpreting the bi-module structure
(3) as a {\it two-sided gauge transformation}$^{[11]}$
in NCGT. 
\\
\indent
In this paper, we restrict ourselves to the
lepton sector and consider only in 
the broken phase, (7), that is, NCQED
of leptons. Hence we encounter only $u$ but {\it both} $u$ {\it and} $u^*$
appear in the gauge transformations.
In later communications, we will consider NCSM
in the lepton and quark sectors.
Although NCQED has been the subject of intensive study
in its own interests,
it is based on the Lorentz-non-covariant algebra
$[{\hat x}^\mu,{\hat x}^\nu]=i\theta^{\mu\nu}$
with constant antisymmetric matrix $\theta^{\mu\nu}$.
Hence it violates Lorentz symmetry.
Quite recently, Carlson, Carone and Zobin$^{[7]}$
constructed a Lorentz-invariant NCGT
by employing DFR(Doplicher-Fredenhagen-Roberts) algebra$^{[12]}$
of non-commutative space-time,
which replaces $\theta^{\mu\nu}$ with an anti-symmetric 
tensor operator ${\hat \theta}^{\mu\nu}$.
According to their formulation the old $\theta^{\mu\nu}$
plays a role as an argument of 4-dimensional covariant fields
as extra 6-dimensional coordinates. 
Consequently, the action contains an integration
over the extra dimensions, too, with unknown weighting function $W(\theta)$.
The $\theta$ in the old version becomes an integration
variable. 
The function $W(\theta)$ is not yet determined from the 
first principle except for normalization
and its evenness.
However, the authors in Ref. 7)
performed detailed perturbational calculation for $2\gamma\to 2\gamma$
scattering to compare with the standard model prediction.
We shall show using a simple NC scalar model 
that the $\theta$ integration with a model choice of $W(\theta)$ gives an {\it invariant} 
damping factor instead of the oscillating one
to the nonplanar self-energy diagram in the one-loop approximation.
The IR-singularity which can not be
discriminated from the $\theta\to 0$ singularity,
may be avoided
if we take UV limit
{\it at the same time} as the commutative limit.
The $\theta$
expansion$^{[14]}$ using Seiberg-Witten map$^{[15]}$
defines QED in the {\it smooth} commutative limit
in the Lagrangian level,
while higher order terms in $\theta$
involve exotic but well-organized derivative interactions$^{[14]}$.
\\
\indent
To reveal another aspects of NCQED relevant to NC bi-module
we recall that the minimal interaction in QED
can be written in two different but equivalent ways,
$$
e{\bar\psi}\gamma^\mu A_\mu\psi=e{\bar\psi}\gamma^\mu\psi A_\mu.
$$
Since one can not freely move the operators on non-commutative space-time,
the above two ways of writing force us to naturally consider the
different gauge fields in NCQED,
one corresponding to $A_\mu$ sandwiched between the spinors
and the other to $A_\mu$ outside the spinors
\footnote{
Since we take trace on operator space,
whether we move $A_\mu$ to the left or to the right
of the spinors
is irrelevant.}.
They are destined to fuse into 
the single photon field in the commutative limit
\footnote{
Two-sided gauge transformation
is familiar in NCGT.
Our assertion is that it
happens to make two independent gauge fields fuse into
a single one in the commutative limit.}.
\\
\indent
By the same token we are motivated to introduce$^{[11]}$
two different charge conjugation transformations in NCQED
since, in QED, we have two different but equivalent
ways of writing the charge conjugation, 
$$
(e{\bar\psi}\gamma^\mu A_\mu\psi)^c=
\left\{
\begin{array}{l}
e{\bar\psi}^c\gamma^\mu A_\mu^c\psi^c,\\
-e\psi^{cT}\gamma^{\mu T} A_\mu^c{\bar\psi}^{cT}.\\
\end{array}
\right.
$$
The latter line is the proper generalization
of the usual charge conjugation $C$, while
the former defines another charge conjugation transformation,
called $C'$. 
\\
\indent
The plan of this paper goes as follows.
We define fields on DFR algebra in the next section,
which largely owes Ref.7).
In section 3, we construct Lorentz-invariant NCQED
to accommodate fermions with the Abelian charges $\pm 2,\pm 1,0$.
As shown in Ref. 7), technically, there is only a slight
modification from the Lorentz-non-invariant NCQED.
Our NCQED possesses two gauge fields which
are each other's $C'$ conjugates.
In the section 4, we shall demonstrate that the $\theta$ integration
actually gives an invariant damping factor in loop-integration 
using a simple NC scalar model in Euclidean metric
and suggest a way to avoid
IR-singularity$^{[13]}$.
We discuss Seiberg-Witten map in our
NCQED characterized by the two gauge fields to be identified
with the single
photon field in the commutative limit in the section 5.
Conclusions are given in the last section.
Appendix A discusses a non-smoothness in the commutative limit of 
the derivative operator for constant $\theta$ algebra.
In the Appendix B a proof will be given on
the correspondence of operator product to Moyal product.

\section{Fields defined on DFR algebra}

\indent
A field on a non-commutative space-time is an operator
in a classical sense. 
It must have a definite
transformation property under the Lorentz group acting on the operator
coordinates ${\hat x}^\mu$. 
If it defines a Lorentz-invariant theory,
the algebra of the operator coordinates must be Lorentz-covariant.
Namely, ${\hat x}'{}^\mu=\Lambda^\mu_{\;\;\nu}{\hat x}^\nu$ 
must satisfy the same algebra as 
the algebra obeyed by ${\hat x}^\mu$ as viewed
in the primed reference frame connected with
the unprimed reference frame by a Lorentz transformation $(\Lambda^\mu_{\;\;\nu})$.
In conformity with this requirement we employ as in Re. 7)
DFR (Doplicher-Fredenhagen-Roberts) algebra$^{[12]}$ spanned by the
hermitian operators ${\hat x}^\mu$ and ${\hat \theta}^{\mu\nu}=-{\hat \theta}^{\nu\mu}$ ($\mu,\nu=0,1,2,3)$
which satisfy the commutation relations
\begin{eqnarray}
[{\hat x}^\mu,{\hat x}^\nu]&=&i{\hat \theta}^{\mu\nu},\;\;\:
[{\hat \theta}^{\mu\nu},{\hat x}^\rho]=
[{\hat \theta}^{\mu\nu},{\hat \theta}^{\rho\sigma}]=0.
\end{eqnarray}
This algebra is Lorentz-covariant, allowing us to 
define operator scalar, spinor, vector and tensor fields
${\hat \varphi}({\hat x},{\hat \theta})$. For instance,
if ${\hat \varphi}({\hat x},{\hat \theta})={\hat A}_\mu({\hat x},{\hat \theta})$
is operator vector field, it transforms as
${{\hat A}'}_\mu({\hat x}',{\hat \theta}')=\Lambda_\mu^{\;\;\nu}{\hat A}_\nu({\hat x},{\hat \theta})$
where ${\hat x}'{}^\mu=\Lambda^\mu_{\;\;\nu}{\hat x}^\nu$ and
${\hat \theta}'{}^{\mu\nu}=\Lambda^\mu_{\;\;\rho}\Lambda^\nu_{\;\;\sigma}
{\hat \theta}^{\rho\sigma}$.
If we replace ${\hat \theta}^{\mu\nu}$ with
$\theta^{\mu\nu}$, 
a real constant antisymmetric matrix,
we write
${\hat \varphi}({\hat x})$ instead of ${\hat \varphi}({\hat x},{\hat \theta})$. Although we may impose the
condition ${\hat \varphi}'({\hat x}')={\hat \varphi}({\hat x})$ to
define operator scalar field,
the algebra spanned by ${\hat x}^\mu$ with constant $\theta$
is no longer Lorentz covariant. Hence Lorentz symmetry is lost
in any theory based on the algebra $[{\hat x}^\mu,{\hat x}^\nu]=i\theta^{\mu\nu}$
for constant $\theta$.
Only in the limit $\theta^{\mu\nu}=0$,
the simultaneous eigenvalues of ${\hat x}^\mu$ can be regarded as
a label of a point in $M_4$ where we define a scalar field by
$\varphi'(x')=\varphi(x)$ with $x'=\Lambda x$ and ${x'}^2=x^2$
\footnote{
This invariance is replaced by
the Lorentz covariance of the algebra spanned by the operator coordinates.}.
However, the commutative limit is discontinuous in this case.
A symptom concerns with the derivative operator, which will be discussed 
in the Appendix A. 
This suggests that
the Lorentz invariance should be maintained from the outset.
In the series of papers we employ Lorentz invariant formulation by
Carlson-Carone-Zobin$^{[7]}$.
\\
\indent
It is well-known that
the field ${\hat \varphi}({\hat x},{\hat \theta})$
is in one to one correspondence with $c$-number field $\varphi(x,\theta)$.
The field $\varphi(x,\theta)$ is obtained by replacing 
the operators ${\hat x}^\mu, {\hat\theta}^{\mu\nu}$
in the Weyl representation
\begin{eqnarray}
{\hat \varphi}({\hat x},{\hat \theta})&=&\frac 1{(2\pi)^4}
\int\!d^4kd^6\sigma{\tilde\varphi}(k,\sigma)e^{ik_\mu{\hat x}^\mu
+i\sigma_{\mu\nu}{\hat\theta}^{\mu\nu}}
\equiv\frac 1{(2\pi)^4}
\int\!d^4kd^6\sigma{\tilde\varphi}(k,\sigma)e^{ik{\hat x}+i\sigma{\hat\theta}},
\end{eqnarray}
with $c$-numbers
$x^\mu, \theta^{\mu\nu}$, respectively, 
\begin{eqnarray}
\varphi(x,\theta)&=&
\frac 1{(2\pi)^4}\int\!d^4kd^6\sigma
{\tilde\varphi}(k,\sigma)e^{ikx+i\sigma\theta}.
\end{eqnarray}
It goes without saying that,
if ${\hat \varphi}({\hat x},{\hat \theta})$ is operator scalar, spinor, vector, and tensor fields,
then $\varphi(x,\theta)$ is also scalar, spinor, vector, and tensor fields,
respectively.
Although it has ten-dimensional arguments,
its transformation property is defined with respect to
the 4-dimensional Lorentz group.
If we put $\theta^{\mu\nu}=0$ and define 
${\tilde\varphi}(k)=\int\!d^6\sigma
{\tilde\varphi}(k,\sigma)$, (10)
gives the usual Fourier transform of a 4-dimensional field $\varphi(x)\equiv
\varphi(x,0)$
with Fourier component ${\tilde\varphi}(k)$.
Consequently, the limit $\theta^{\mu\nu}\to0$
corresponds to the commutative limit
\footnote{
This corresponds to taking the limit $[{\hat x}^\mu,{\hat x}^\nu]\to 0$,
since the commutator $[{\hat x}^\mu,{\hat x}^\nu]$ is proportional to
the fundamental length squared, $a^2$, as in Snyder's algebra$^{[16]}$ 
and the commutative limit sends $a$ to 0.
However, this limit is much mild than the usual one,
$[{\hat x}^\mu,{\hat x}^\nu]=i\theta^{\mu\nu}$, at $\theta^{\mu\nu}\to 0$
and an embarrassment encountered in defining the derivative operator 
in the latter case is avoided. We shall discuss this point in the Appendix A.
Moreover, we can build 4-dimensional field theory
solely based on DFR algebra. Define field ${\hat \varphi}({\hat x},{\hat \theta})$ on DFR algebra
and take the limit $\theta^{\mu\nu}\to0$. 
Quantization is to be performed on the $c$-number field $\varphi(x)=\varphi(x,\theta=0)$.
(We shall discuss this point on later occasion.)
In this sense, the commutative limit
is {\it smooth}.}.
The inverse Fourier transform is given by
\begin{eqnarray}
{\tilde\varphi}(k,\sigma)&=&\frac 1{(2\pi)^6}
\int\!d^4xd^6\theta
\varphi(x,\theta)e^{-ikx-i\sigma\theta}.
\end{eqnarray}
\indent
Since the translation ${\hat x}^\mu\to {\hat x}^{\mu}+a^\mu{\hat 1}$ for
any $c$-number $a^\mu$ with ${\hat 1}$ being the unit operator
is an automorphism of the DFR algebra, we can define the operator
\begin{eqnarray}
{\hat \varphi}({\hat x}+a{\hat 1},{\hat \theta})=\frac 1{(2\pi)^4}
\int\!d^4xd^6\sigma{\tilde\varphi}(k,\sigma)
e^{ik_\mu({\hat x}^\mu+a^\mu{\hat 1})+i\sigma_{\mu\nu}{\hat\theta}^{\mu\nu}},
\end{eqnarray}
from which the derivative of an operator is defined$^{[12]}$ by
\begin{eqnarray}
\partial_\mu{\hat \varphi}({\hat x},{\hat \theta})&=&
\frac{\partial}{\partial a^\mu}{\hat \varphi}({\hat x}+a{\hat 1},{\hat \theta})|_{a=0}\nonumber\\[2mm]
&=&\frac 1{(2\pi)^{4}}
\int\!d^4kd^6\sigma ik_\mu{\tilde\varphi}(k,\sigma)
e^{ik{\hat x}+i\sigma{\hat\theta}}.
\end{eqnarray}
We also define the trace$^{[7]}$
\footnote{
If ${\hat \varphi}$ is a matrix, matrix trace is also implied.}
\begin{eqnarray}
{\rm tr}\;{\hat \varphi}({\hat x},{\hat \theta})
&=&\frac 1{(2\pi)^6}\int\!d^6\sigma{\tilde\varphi}(0,\sigma){\tilde W}(\sigma)
=\int\!d^4xd^6\theta\varphi(x,\theta)W(\theta),
\label{eqn:2-7}
\end{eqnarray}
where
\begin{eqnarray}
W(\theta)
&=&\frac 1{(2\pi)^6}\int\!d^6\sigma {\tilde W}(\sigma)e^{-i\sigma\theta},
\label{eqn:2-8}
\end{eqnarray}
with the normalization
\begin{eqnarray}
\int\!d^6\theta W(\theta)=1.
\label{eqn:2-9}
\end{eqnarray}
It is clear that the function $W(\theta)$ has dimensions,
$[L^{-12}]$. Taking the trace of the first equation
of (8) we get tr$\;{\hat \theta}^{\mu\nu}=0$ so that
\begin{eqnarray}
\int\!d^6\theta W(\theta)\theta^{\mu\nu}=0,
\end{eqnarray}
since Weyl
symbol corresponding to the operator ${\hat \theta}^{\mu\nu}$
is $\theta^{\mu\nu}$. Thus $W(\theta)$ is an even function
\footnote
{The authors in Ref. 7) are led to the condition $W(\theta)=W(-\theta)$
from Lorentz invariance.}.
The commutative limit corresponds to
\begin{eqnarray}
W(\theta)=W^{(0)}(\theta)\equiv\delta^6(\theta)\equiv \delta(\theta^{01})
\delta(\theta^{02})\delta(\theta^{03})
\delta(\theta^{12})\delta(\theta^{23})\delta(\theta^{31}).
\end{eqnarray}
This has a correct dimension, $[L^{-12}]$, since
$\theta$ has dimensions of length squared.
\\
\indent
The Weyl symbol of the operator product is given by the Moyal product.
To see this put 
\begin{eqnarray}
{\hat \varphi}_1({\hat x},{\hat \theta}){\hat \varphi}_2({\hat x},{\hat \theta})&=&
\frac 1{(2\pi)^4}
\int\!d^4kd^6\sigma {\tilde \varphi}_{12}(k,\sigma)e^{ik{\hat x}+i\sigma{\hat\theta}}.
\end{eqnarray}
Then we can show that
\begin{eqnarray}
\varphi_{12}(x,\theta)&=&
\frac 1{(2\pi)^4}
\int\!d^4kd^6\sigma{\tilde \varphi}_{12}(k,\sigma)e^{i(kx+\sigma\theta)}\nonumber\\[2mm]
&=&e^{\frac i2\theta^{\mu\nu}\frac{\partial}{\partial x^\mu}
\frac{\partial}{\partial y^\nu}}
\varphi_1(x,\theta)\varphi_2(y,\theta)|_{x=y}\nonumber\\[2mm]
&\equiv&\varphi_1(x,\theta)*\varphi_2(x,\theta).
\end{eqnarray}
The proof will be given in the Appendix B.
Namely,
if ${\hat \theta}^{\mu\nu}$ belongs to the center of the algebra,
the product of operators corresponds to the Moyal product
as if ${\hat \theta}^{\mu\nu}$ is a $c$-number $\theta^{\mu\nu}$.
Only difference lies in the additional dependence of
the Weyl symbol on $\theta$, $\varphi(x,\theta)$.
This is the most important observation by Carlson-Carone-Zobin$^{[7]}$.
\\
\indent
By definition (20) we have
\begin{equation}
\int\!d^4x\varphi_1(x,\theta)*\varphi_2(x,\theta)
=\int\!d^4x\varphi_1(x,\theta)\varphi_2(x,\theta)
=\int\!d^4x\varphi_2(x,\theta)*\varphi_1(x,\theta),
\end{equation}
so that it follows
\begin{eqnarray}
{\rm tr}\;{\hat \varphi}_1({\hat x},{\hat \theta}){\hat \varphi}_2({\hat x},{\hat \theta})
&=&\int\!d^4xd^6\theta W(\theta)\varphi_1(x,\theta)\varphi_2(x,\theta)
=\int\!d^4xd^6\theta W(\theta)\varphi_2(x,\theta)\varphi_1(x,\theta)\nonumber\\[2mm]
&=&{\rm tr}\;{\hat \varphi}_2({\hat x},{\hat \theta}){\hat \varphi}_1({\hat x},{\hat \theta}).
\end{eqnarray}
\indent
Moyal product of three Weyl symbols is similarly defined.
\begin{eqnarray}
{\hat \varphi}_1({\hat x},{\hat \theta}){\hat \varphi}_2({\hat x},{\hat \theta}){\hat \varphi}_3({\hat x},{\hat \theta})
&=&\frac 1{(2\pi)^4}
\int\!d^4kd^6\sigma{\tilde \varphi}_{123}(k,\sigma)e^{ik{\hat x}+i\sigma{\hat\theta}}\nonumber\\[2mm]
\varphi_{123}(x,\theta)
&=&\frac 1{(2\pi)^4}
\int\!d^4kd^6\sigma{\tilde \varphi}_{123}(k,\sigma)e^{i(kx+\sigma\theta)}\nonumber\\[2mm]
&=&\varphi_1(x,\theta)*\varphi_2(x,\theta)*\varphi_3(x,\theta).
\end{eqnarray}
The associativity is proven from that of operators.
Using the cyclic property of the trace 
we can show the cyclic property of Moyal products under integration
\begin{eqnarray}
\int\!d^4x\varphi_1(x,\theta)*\varphi_2(x,\theta)*\varphi_3(x,\theta)&=&
\int\!d^4x\varphi_2(x,\theta)*\varphi_3(x,\theta)*\varphi_1(x,\theta)\nonumber\\[2mm]
&=&\int\!d^4x\varphi_3(x,\theta)*\varphi_1(x,\theta)*\varphi_2(x,\theta),
\end{eqnarray}
where we have omitted integration $\int\!d^6\theta W(\theta)$.

\section{Lorentz-invariant NCQED}

\indent
Let us now consider Lorentz-invariant NCQED for fermion. 
Since fields on DFR algebra (8) only demands additional 
dependence on the variable $\theta$
as compared with those defined on the Lorentz-non-covariant algebra
$[{\hat x}^\mu,{\hat x}^\nu]=i\theta^{\mu\nu}$, Lorentz-invariant NCQED
closely follows from the Lorentz-non-invariant NCQED. 
For the reason explained in the Introduction,
we employ NCQED as detailed in Ref.11).
Thus we consider the following gauge transformations for 8 spinors
\begin{eqnarray}
\chi_1(x,\theta)&\to& ^{\hat g}\chi_1(x,\theta)=U(x,\theta)*\chi_1(x,\theta)*U^{\dag}(x,\theta),\nonumber\\[2mm]
\chi_2(x,\theta)&\to& ^{\hat g}\chi_2(x,\theta)=U^{\dag}(x,-\theta)*\chi_2(x,\theta)*U(x,-\theta),\nonumber\\[2mm]
\psi_1(x,\theta)&\to& ^{\hat g}\psi_1(x,\theta)=U(x,\theta)*\psi_1(x,\theta),\nonumber\\[2mm]
\psi_2(x,\theta)&\to& ^{\hat g}\psi_2(x,\theta)=\psi_2(x,\theta)*U(x,-\theta),\nonumber\\[2mm]
\psi_3(x,\theta)&\to& ^{\hat g}\psi_3(x,\theta)=U^{\dag}(x,-\theta)*\psi_3(x,\theta),\nonumber\\[2mm]
\psi_4(x,\theta)&\to& ^{\hat g}\psi_4(x,\theta)=\psi_4(x,\theta)*U^{\dag}(x,\theta),\nonumber\\[2mm]
\psi_5(x,\theta)&\to& ^{\hat g}\psi_5(x,\theta)=U(x,\theta)*\psi_5(x)*U(x,-\theta),\nonumber\\[2mm]
\psi_6(x,\theta)&\to& ^{\hat g}\psi_6(x,\theta)=U^{\dag}(x,-\theta)*\psi_6(x,\theta)*U^{\dag}(x,\theta),
\end{eqnarray}
where the gauge parameter is assumed to be $*$ unitary, 
\begin{eqnarray}
U(x,\theta)*U^{\dag}(x,\theta)&=&U^{\dag}(x,\theta)*U(x,\theta)=1,\nonumber\\[2mm]
U(x,\theta)&=&(e^{i\alpha(x,\theta)})_*\nonumber\\[2mm]
&\equiv&
1+i\alpha(x,\theta)+\frac 1{2!}(i\alpha(x,\theta))*(i\alpha(x,\theta))
+\cdots.
\end{eqnarray}
The product of $U(x,-\theta)$ is defined by
\begin{eqnarray}
U_1(x,-\theta){\bar *}U_2(x,-\theta)\equiv
U_2(x,-\theta)*U_1(x,-\theta),
\end{eqnarray}
so that the group property with respect to $\ast$ product is retained,
\begin{eqnarray}
U_1^{\dag}(x,-\theta){\bar *}U_2^{\dag}(x,-\theta)\equiv
U_2^{\dag}(x,-\theta)*U_1^{\dag}(x,-\theta)
=(U_1(x,-\theta)*U_2(x,-\theta))^{\dag}.
\end{eqnarray}
In Ref. 11) we did not consider the $\theta$ dependence
of the gauge parameter, although we implicitly assumed
the $\theta$ dependence of the fields including the gauge fields.
(See the section 5.)
\\
\indent
In the commutative limit, we have only 5 spinors
all of which receive
$U(1)$ gauge transformation
\begin{eqnarray}
\psi(x)\to ^g\!\!\psi(x)=U(x)\psi(x),\quad U(x)=e^{iQ\alpha(x)},
\end{eqnarray}
where $\alpha(x)=\alpha(x,0)$ and the charge $Q$ is determined as follows.\\[2mm]
1)$\;\;$
The set $\{\chi_1, \psi_1,\psi_4\}$ couples to
the gauge field transforming like
\begin{eqnarray}
A_\mu(x,\theta)&\to& ^{\hat g}\!\!A_\mu(x,\theta)=U(x,\theta)*A_\mu(x,\theta)*U^{\dag}(x,\theta)
+\frac{2i}e
           U(x,\theta)*\partial_\mu U^{\dag}(x,\theta).
\label{eqn:3-6}
\end{eqnarray}
As a consequence, $Q(\chi_1)=0, Q(\psi_1)=+1$ and $Q(\psi_4)=-1$ in units of $e/2$. 
This is the charge quantization in NCQED$^{[10]}$. \\[2mm]
2)$\;\;$
On the other hand, the set $\{\chi_2, \psi_2,\psi_3\}$ interacts
with another gauge field with different
transformation property
\begin{equation}
{A'}_\mu(x,\theta)\to ^{\hat g}\!\!{A'}_\mu(x,\theta)=U^{\dag}(x,-\theta)*{A'}_\mu(x,\theta)*U(x,-\theta)
+\frac{2i}e
           U^{\dag}(x-\theta)*\partial_\mu U(x-\theta).
\end{equation}
Hence, $Q(\chi_2)=0, Q(\psi_2)=+1$ and $Q(\psi_3)=-1$ in units of $e/2$.
This is also the charge quantization in NCQED. \\[2mm]
3)$\;\;$
Apparently, $\{\psi_5,\psi_6\}$ with $Q(\psi_5)=+1$ and
$Q(\psi_6)=-1$ in units of $e$ couple to both gauge fields.
In the commutative limit, we may put $\psi_1=\psi_2$, $\psi_3=\psi_4$ 
(and $\chi_1=\chi_2$)
so that
the two gauge fields should become identical up to sign.
That is, the photon field $A_\mu(x)$ is given by
\begin{equation}
A_\mu(x)=A_\mu(x,0)=-{A'}_\mu(x,0).
\end{equation}
In fact, inverting the sign of $\theta$ in (30)
and comparing the result with (31), we can put
\begin{equation}
{A'}_\mu(x,\theta)=-A_\mu(x,-\theta),
\end{equation}
since $\Delta_\mu\equiv
{A'}_\mu(x,\theta)+A_\mu(x,-\theta)$ is subject to homogeneous transformation
and can be put zero in a gauge-invariant way
\footnote{
In general one can only say that $\Delta_\mu\to 0$ at $\theta\to 0$.}.
\\
\indent
In the commutative limit, we have the single photon field which couples
to four spinors, $\psi_1=\psi_2, \psi_3=\psi_4, \psi_5$ and $\psi_6$, only. In other words,
if we define QED as a limiting theory of Lorentz-invariant NCQED,
we are allowed to have only four spinors with
charges $\pm 2,\pm 1$ in units of $e/2$ and one neutral spinor.
We shall give very plausible argument in favor of the spinors
$(\chi_1,\psi_6)$ 
to represent leptons in Nature in later
publications where
we obtain our NCQED
from a spontaneously broken gauge theory of Lorentz-invariant non-commutative Weinberg-Salam model
(NCWS). ((1)
implies two kinds of spinors $\chi_1$ and $\psi_6$ to describe leptons.
See the footnote on p.4.)
This is why we employed NCQED in Ref.11), which considers
all possible spinors. In other references, such spinors like $\psi_5$
or $\psi_6$, which will be responsible for the charged leptons
considering the bi-module structure (1)
literally, are not introduced.
\\
\indent
We now list all gauge couplings for 8 spinors
\begin{eqnarray}
&&\frac e2{\bar\chi}_1(x,\theta)*\gamma^\mu(A_\mu(x,\theta)*\chi_1(x,\theta)
-\chi_1(x,\theta)*A_\mu(x,\theta)),\nonumber\\[2mm]
&&\frac e2{\bar\chi}_2(x,\theta)*\gamma^\mu({A'}_\mu(x,\theta)*\chi_2(x,\theta)
-\chi_2(x,\theta)*{A'}_\mu(x,\theta)),\nonumber\\[2mm]
&&\frac e2{\bar\psi}_1(x,\theta)*\gamma^\mu A_\mu(x,\theta)*\psi_1(x,\theta),\nonumber\\[2mm]
&&
-\frac e2{\bar\psi}_2(x,\theta)*\gamma^\mu\psi_2(x,\theta)*{A'}_\mu(x,\theta),\nonumber\\[2mm]
&&\frac e2{\bar\psi}_3(x,\theta)*\gamma^\mu {A'}_\mu(x,\theta)*\psi_3(x,\theta),\nonumber\\[2mm]
&&
-\frac e2{\bar\psi}_4(x,\theta)*\gamma^\mu\psi_4(x,\theta)*A_\mu(x,\theta),\nonumber\\[2mm]
&&\frac e2{\bar\psi}_5(x,\theta)*\gamma^\mu (A_\mu(x,\theta)*\psi_5(x,\theta)
-\psi_5(x,\theta)*{A'}_\mu(x,\theta)),\nonumber\\[2mm]
&&\frac e2{\bar\psi}_6(x,\theta)*\gamma^\mu ({A'}_\mu(x,\theta)*\psi_6(x,\theta)
-\psi_6(x,\theta)*A_\mu(x,\theta)).
\end{eqnarray}
They all conserve $C$ if we define
\begin{eqnarray}
{\cal C}\psi(x,\theta){\cal C}^{-1}&\equiv&\psi^c(x,\theta)=C{\bar\psi}^T(x,\theta),\nonumber\\[2mm]
{\cal C}{\bar\psi}(x,\theta){\cal C}^{-1}&\equiv&{\bar\psi}^c(x,\theta)=-\psi^T(x,\theta) C^{-1},\nonumber\\[2mm]
{\cal C} A_\mu(x,\theta){\cal C}^{-1}&\equiv& A_\mu^c(x,\theta)=-A_\mu(x,\theta),\nonumber\\[2mm]
{\cal C} {A'}_\mu(x,\theta){\cal C}^{-1}&\equiv& {A'}_\mu^c(x,\theta)=-{A'}_\mu(x,\theta),
\end{eqnarray}
where $C$ is the charge conjugation matrix satisfying
$C^{-1}\gamma^\mu C=-\gamma^{\mu T}$,
provided ${\cal C}$ reverses the order of the operators (the order of 
Moyal product).
For instance, the gauge coupling for the spinor $\chi_1$ upon integration
is invariant under $C$
\begin{eqnarray}
&&\!\!\!\!\!\!\!\!\!\!
-\frac e2\int\!d^4xd^6\theta W(\theta)[\chi_1^{cT}(x,\theta)*\gamma^{\mu T}
(A_\mu^c(x,\theta)*{\bar\chi}_1^{cT}(x,\theta)
-{\bar\chi}_1^{cT}(x,\theta)*A_\mu^c(x,\theta))]\nonumber\\[2mm]
&&\!\!\!\!\!\!\!\!\!\!
=\frac e2\int\!d^4xd^6\theta W(\theta)
[{\bar\chi}_1(x,\theta)*\gamma^\mu(A_\mu(x,\theta)*\chi_1(x,\theta)
-\chi_1(x,\theta)*A_\mu(x,\theta))].
\end{eqnarray}
On the other hand, the gauge coupling for the spinor $\psi_5$
\begin{eqnarray}
&&\frac e2\int\!d^4xd^6\theta W(\theta)[{\bar\psi}_5(x,\theta)
*\gamma^\mu A_\mu(x,\theta)*\psi_5(x,\theta)\nonumber\\[2mm]
&&\qquad\qquad\qquad\qquad+
{\bar\psi}_5^c(x,\theta)*\gamma^\mu {A'}_\mu(x,\theta)*\psi_5^c(x,\theta)],
\end{eqnarray}
is invariant under the transformation ($\psi_5\to\psi$)
\begin{eqnarray}
{\cal C}'\psi(x,\theta){{\cal C}'}^{-1}&=&\psi^c(x,\theta)=C{\bar\psi}^T(x,\theta),\nonumber\\[2mm]
{\cal C}'{\bar\psi}(x,\theta){{\cal C}'}^{-1}&=&{\bar\psi}^c(x,\theta)=-\psi^T(x,\theta)C^{-1},\nonumber\\[2mm]
{\cal C}'A_\mu(x,\theta){{\cal C}'}^{-1}&=&A'_\mu(x,\theta),\;\;\;\;
{\cal C}'A'_\mu(x,\theta){{\cal C}'}^{-1}=A_\mu(x,\theta),
\end{eqnarray}
provided the order of the operators (the order of 
Moyal product) are {\it not} reversed.
Similarly for the spinor $\psi_6$.
We call the transformation 
\begin{eqnarray}
     &&\psi(x,\theta)\leftrightarrow \psi^c(x,\theta),\;\;\;
     {\bar \psi}(x,\theta)\leftrightarrow {\bar \psi}^c(x,\theta),\;\;\;
     A_\mu(x,\theta)\leftrightarrow {A'}_\mu(x,\theta),
\end{eqnarray}
for the spinors $\psi=\psi_5, \psi_6$, $C'$ transformation$^{[11]}$
provided no reversal of operators (the order of the Moyal product)
is made. 
Then we may say that
${A'}_\mu(x,\theta)$ is $C'$ conjugate of $A_\mu(x,\theta)$.
There are two charge conjugations $C$ and $C'$, the difference
being the reversal and unreversal of the operators, respectively.
In the commutative limit, the order of the operators become irrelevant
so that $C'\to C$ at $\theta=0$. 
Hence follows (32) by definition.
\\
\indent
To conclude this section we define $C'$-invariant NC Maxwell action by
\begin{eqnarray}
{\hat S}_M&=&-\frac 18\int\!d^4xd^6\theta W(\theta)(
F_{\mu\nu}(x,\theta)*F^{\mu\nu}(x,\theta)
+{F'}_{\mu\nu}(x,\theta)*{F'}^{\mu\nu}(x,\theta)),\nonumber\\[2mm]
F_{\mu\nu}(x,\theta)&=&\partial_\mu A_\nu(x,\theta)-\partial_\nu A_\mu(x,\theta)\nonumber\\[2mm]
      &&-\frac{ie}2(A_\mu(x,\theta)*A_\nu(x,\theta)-A_\nu(x,\theta)*A_\mu(x,\theta)),\nonumber\\[2mm]
{F'}_{\mu\nu}(x,\theta)&=&F_{\mu\nu}|_{A_\mu(x,\theta)\to {A'}_\mu(x,\theta)}.
\end{eqnarray}
Each term $F^2$ and ${F'}^2$ are separately $C$-invariant,
\begin{eqnarray}
F_{\mu\nu}^c(x,\theta)&=&\partial_\mu(-A_\nu(x,\theta))-\partial_\nu(-A_\mu(x,\theta))\nonumber\\[2mm]
      &&-\frac {ie}2[(-A_\nu(x,\theta))*(-A_\mu(x,\theta))
      -(-A_\mu(x,\theta))*(-A_\nu(x,\theta))]\nonumber\\[2mm]
      &=&-F_{\mu\nu}(x,\theta),
\end{eqnarray}
and similarly for ${F'}_{\mu\nu}(x,\theta)$.
In the subsequent paper we derive (40)
from Lorentz-invariant NCWS model.
Due to the condition (33)
we easily see that
\begin{eqnarray}
{F'}_{\mu\nu}(x,\theta)&=&-F_{\mu\nu}(x,-\theta),\nonumber\\[2mm]
A_\mu(x,\theta)*A_\nu(x,\theta)|_{\theta\to -\theta}&=&
A_\nu(x,-\theta)*A_\mu(x,-\theta).
\end{eqnarray}
That is, $C'$-invariant NC Maxwell Lagrangian is even in $\theta$.
\section{An invariant damping factor}
The original motivation of
introducing quantized space-time$^{[16]}$
was to remove UV divergence troubles in quantum field theory
by replacing point-like interactions of elementary
particles with specific Lorentz-invariant nonlocal interactions.
On the other hand, Filk$^{[17]}$ showed that,
based on the algebra $[{\hat x}^\mu,{\hat x}^\nu]=i\theta^{\mu\nu}$,
the nonplanar diagrams receive an oscillating
damping factor due to the space-time non-commutativity,
whereas UV divergence of the planar diagrams remain unchanged.
We would like to emphasize that the oscillating damping factor
violates Lorentz invariance. In fact,
the authors in Re.13) observed
in non-commutative scalar model that
1PI two-point function in the one-loop approximation 
explicitly violates
Lorentz symmetry,
and exhibited
a singular behavior at $\theta\to 0$ limit after
loop integration. Since $\theta$
appears always through the combination
$\theta^{\mu\nu}q_\nu$ in the oscillating damping factor,
where $q_\nu$
is the external momentum,
this singular behavior necessarily implies
IR-singularity$^{[13]}$ of the amplitude
at $\theta^{\mu\nu}q_\nu\to 0$.
Hayakawa$^{[10]}$ independently pointed out an explicit
violation of Lorentz symmetry
in the one-loop photon propagator in NCQED and investigated
IR behavior together with the  $\theta\to 0$
singularity in relation with UV divergence.
It would be interesting to see what
happens on the oscillating damping factor
if we employ the Lorentz-invariant
formulation.
\\
\indent
In order to study this problem we consider
a simpler example, namely, NC scalar $\lambda\phi^4$-theory
in Euclidean metric.
In the old Lorentz-non-invariant version it is defined by
\begin{eqnarray}
{\hat S}=\int\!d^4x[
\frac 12(\partial_\mu\phi(x)\partial^\mu\phi(x)+m^2\phi^2(x))
+\frac \lambda{4!}\phi(x)*\phi(x)*\phi(x)*\phi(x)].
\end{eqnarray}
The proper self-energy part in the one-loop approximation
is given by$^{[13]}$
\begin{eqnarray}
\Sigma^{(1)}_{\rm pl}&=&
\frac \lambda{3(2\pi)^4}\int\!d^4k\frac 1{k^2+m^2},\nonumber\\[2mm]
\Sigma^{(1)}_{\rm npl}&=&
\frac \lambda{6(2\pi)^4}\int\!d^4k\frac {e^{ik_\mu\theta^{\mu\nu}p_\nu}}{k^2+m^2},
\end{eqnarray}
for planar and nonplanar diagrams, respectively.
It is well-known$^{[17]}$ that the nonplanar diagram is UV finite
due to the oscillating factor, $e^{ik_\mu\theta^{\mu\nu}p_\nu}$,
where $p_\nu$ is the external momentum.
This oscillating factor comes from the nonlocal interaction
concealed in the star product of (43).
Using the Schwinger representation, $\frac 1{k^2+m^2}=
\int_0^\infty\!d\alpha e^{-\alpha(k^2+m^2)}$ and regularizing by
multiplication of the factor $e^{-1/\alpha \Lambda^2}$, we get
\begin{eqnarray}
\Sigma^{(1)}_{\rm pl}&=&
\frac \lambda{48\pi^2}\int_0^\infty\!\frac {d\alpha}{\alpha^2}e^{-\frac 1{\alpha \Lambda^2}
-\alpha m^2}=\frac \lambda{48\pi^2}[\Lambda^2-m^2\ln{\frac{\Lambda^2}{m^2}}
+{\cal O}(1)],\nonumber\\[2mm]
\Sigma^{(1)}_{\rm npl}&=&
\frac \lambda{96\pi^2}\int_0^\infty\!\frac {d\alpha}{\alpha^2}e^{-\frac 1{\alpha \Lambda_{eff}^2}
-\alpha m^2}=\frac \lambda{96\pi^2}[\Lambda_{eff}^2-m^2\ln{\frac{\Lambda_{eff}^2}{m^2}}
+{\cal O}(1)],
\end{eqnarray}
where we have defined
\begin{eqnarray}
\frac 1{\Lambda_{eff}^2}&=&\frac 1{\Lambda^2}+\frac {{\tilde p}^2}4,\nonumber\\[2mm]
{\tilde p}^2&=&\theta^{\mu\nu}p_\nu \theta_{\mu\rho}p^\rho.
\end{eqnarray}
Here, $\theta^{\mu\nu}=\theta_{\mu\nu}$
\footnote
{In Minkowski space-time,
the indices of $\theta^{\mu\nu}$ are lowered by the Lorentz metric,
$\theta_{\mu\nu}=g_{\mu\rho}g_{\nu\sigma}\theta^{\rho\sigma}$,
$(g_{\mu\nu})$=diag($+1,-1,-1,-1)$.}.
It is clear that the nonplanar diagram is UV-finite as far as ${\tilde p}^2\not=0$.
However, it is IR-singular at ${\tilde p}^2\to 0$
if we first let
$\Lambda^2\to\infty$. In order to avoid the IR-singularity
in the limit $\Lambda^2\to\infty$,
we are tempted to take the limit $\theta\to 0$ {\it simultaneously}.
Thus we have instead of the last expression of (45),
\begin{eqnarray}
\Sigma^{(1)}_{\rm npl}&=&
\frac \lambda{96\pi^2}\int_0^\infty\!\frac {d\alpha}{\alpha^2}e^{-\frac 1{\alpha \Lambda^2}
-\alpha m^2}[1-\frac {{\tilde p}^2}{4\alpha}+
\frac{{\tilde p}^4}{2!(4\alpha)^2}+\cdots]\nonumber\\[2mm]
&=&\frac \lambda{96\pi^2}[2(\Lambda^2m^2)^{1/2}K_1(2\sqrt{m^2/\Lambda^2}\,)
-\frac{{\tilde p}^2}42(\Lambda^2m^2)K_2(2\sqrt{m^2/\Lambda^2}\,)\nonumber\\[2mm]
&&+\frac{{\tilde p}^4}{32}2(\Lambda^2m^2)^{3/2}K_3(2\sqrt{m^2/\Lambda^2}\,)+\cdots],
\end{eqnarray}
where $K_n$ is the modified Bessel function.
If we put
\begin{eqnarray}
\theta^{\mu\nu}=a^2{\bar\theta}^{\mu\nu},
\end{eqnarray}
where $a$ is the fundamental length 
with ${\bar\theta}^{\mu\nu}$
being dimensionless,
then the commutative limit corresponds to $a\to 0$.
Near IR region which is indistinguishable from the commutative limit
in the present model, 
we propose to take UV limit
in the following way,
\begin{eqnarray}
\Lambda^2&\to&\infty,\nonumber\\[2mm]
a^2&\to& 0,\nonumber\\[2mm]
\Lambda^2 a^2&:&{\rm fixed}.
\end{eqnarray}
Then the first term in the bracket $[\;]$ of (47)
diverges quadratically as usual,
the second becomes constant of the order $a^4\Lambda^4$,
the third vanishes because it behaves like $a^8\Lambda^6$
and the rest follows the same fate as the third.
It should be noted, however, that
the drawback of the above argument is the apparent loss of the
Lorentz invariance (here, Euclidean symmetry). Our
argument may work only in the case of $\Sigma^{(1)}_{npl}$
being a function of the invariant, $p^2$.
\\
\indent
This shortcoming is remedied by Carlson-Carone-Zobin formulation$^{[7]}$
of the model,
\begin{eqnarray}
{\hat S}=\int\!d^4xd^6\theta W(\theta)[
\frac 12(\partial_\mu\phi(x)\partial^\mu\phi(x)+m^2\phi^2(x))
+\frac \lambda{4!}\phi(x)*\phi(x)*\phi(x)*\phi(x)].
\end{eqnarray}
In this new Lorentz-invariant (here, Euclidean-invariant)
version we have
\begin{eqnarray}
\Sigma^{(1)}_{\rm pl}&=&
\frac \lambda{3(2\pi)^4}\int\!d^4k\frac 1{k^2+m^2},\nonumber\\[2mm]
\Sigma^{(1)}_{\rm npl}&=&
\frac \lambda{6(2\pi)^4}\int\!d^4kd^6\theta W(\theta)
\frac {e^{ik_\mu\theta^{\mu\nu}p_\nu}}{k^2+m^2},
\end{eqnarray}
where we have used the normalization (16)
to obtain the same expression of $\Sigma^{(1)}_{\rm pl}$ as before.
That is, the planar diagram is no different. The $\theta$-integration
in the nonplanar diagram,
\begin{eqnarray}
I=\int\!d^6\theta W(\theta)e^{ik_\mu\theta^{\mu\nu}p_\nu}
\end{eqnarray}
can be performed as follows.
Assuming
\begin{eqnarray}
W(\theta)=a^{-12}w({\bar\theta}),
\end{eqnarray}
the integral becomes
\begin{eqnarray}
I=\int\!d^6{\bar\theta}w({\bar\theta})e^{ia^2k_\mu{\bar\theta^{\mu\nu}}p_\nu}.
\end{eqnarray}
To proceed further we have to choose the functional form of $w({\bar\theta})=w(-{\bar\theta})$
with $\int\!d^6{\bar\theta}w({\bar\theta})=1$.
There is no guiding principle to determine it.
In what follows, we put for computational purpose only
\begin{eqnarray}
w({\bar\theta})=\frac 1{\pi^3}
e^{-({\bar\theta}^{01})^2-({\bar\theta}^{02})^2-({\bar\theta}^{03})^2-({\bar\theta}^{12})^2
-({\bar\theta}^{23})^2-({\bar\theta}^{31})^2}.
\end{eqnarray}
It is then easy to obtain
\footnote
{It is to be noted that to make $w({\bar\theta})$
Lorentz-invariant the exponent must be at least quartic in ${\bar\theta}$
since ${\bar\theta}^{\mu\nu}{\bar\theta}_{\mu\nu}$ is indefinite.
However, $({\bar\theta}^{\mu\nu}{\bar\theta}_{\mu\nu})^2$ is
positive definite and we may use the formula,
$$
\int\!dxe^{-x^4}=\frac 14\Gamma(\frac 14),
$$
to maintain the normalization.}
\begin{eqnarray}
I=e^{-a^4[k^2p^2-(k\cdot p)^2]/4},
\end{eqnarray}
where $k^2=k_0^2+k_1^2+k_2^2+k_3^2$ (we still use the index 0 instead of 4).
Thus the integral $I$ works as an {\it invariant}
damping factor for the nonplanar diagram as far as
$a$ {\it and} $p$ do not vanish
\footnote
{If we insist to Minkowski space-time formulation, 
we would get similar Lorentz-invariant damping factor for
the nonplanar diagram.
By the way, the oscillating factor is absent if
only the projection of $p$
onto the non-commutative subspace vanishes, but our damping
factor becomes unity only if all components of $p$
vanish.}.
Inserting this result into new $\Sigma^{(1)}_{\rm npl}$ of (51),
we get Lorentz (here, Euclidean) invariant result
\begin{eqnarray}
\Sigma^{(1)}_{\rm npl}&=&
\frac \lambda{6(2\pi)^4}\int_0^\infty\!d\alpha e^{-\frac 1{\alpha \Lambda^2}
-\alpha m^2}\int\!d^4ke^{-\alpha k^2-a^4[k^2p^2-(k\cdot p)^2]/4}\nonumber\\[2mm]
&=&\frac \lambda{96\pi^2}\int_0^\infty\!\frac{d\alpha}
{\sqrt{\alpha}(\alpha+a^4 p^2/4)^{3/2}}e^{-\frac 1{\alpha \Lambda^2}
-\alpha m^2},
\end{eqnarray}
which is UV finite unless $a^4p^2=0$.
It is, however, IR-singular, $\Sigma^{(1)}_{\rm npl}\to \frac 8{a^4p^2}
\times\frac \lambda{96\pi^2}$ 
at $p^2\to 0$,
if we first let $\Lambda^2\to\infty$.
This is the same phenomenon as before$^{[13]}$
except that our result is Lorentz (here, Euclidean) invariant.
To avoid this IR-singularity we take the UV limit
as defined above.
Expanding the exponential containing the small parameter $a^4$
in new $\Sigma^{(1)}_{\rm npl}$ and carrying out the $k$-integration, 
we obtain
\begin{eqnarray}
\Sigma^{(1)}_{\rm npl}
&=&\frac \lambda{96\pi^2}[2\sqrt{m^2\Lambda^2}K_1(2\sqrt{m^2/\Lambda^2}\,)
-\frac 38a^4p^22(m^2\Lambda^2)K_2(2\sqrt{m^2/\Lambda^2}\,)\nonumber\\[2mm]
&&+\frac{15}{128}a^8p^42(m^2\Lambda^2)^{3/2}K_3(2\sqrt{m^2/\Lambda^2}\,)+\cdots].
\end{eqnarray}
Our UV limit reproduces the well-known quadratic divergence
from the first term in the above expansion
in addition to a constant term from the second term.
The third and the rest vanish in our UV limit.
We may of course retain the contributions of order 
$(a^4p^2)^n(\Lambda^2)^{n+1}, n=0,1,2,\cdots$,
(including logs)
in the above $(n+1)$-st term without strictly neglecting
the third and the rest. Our argument to bypass
the IR-singularity is to take the limit $a^2\to 0$
{\it at the same time} as the limit $\Lambda^2\to \infty$
so that the above expansion is effectively truncated.
In this way we have successfully evaded the IR-singularity
in an {\it invariant} way. Outside the region where
(51) is used, $\Sigma^{(1)}_{\rm npl}$ is UV finite.
\\
\indent
We expect that a similar
mechanism works in higher order loops and also in NCQED.

\section{Seiberg-Witten map in NCQED}

In order to investigate
the commutative limit of NCWS for leptons,
we have previously$^{[11]}$proposed $\theta$-expansion,
\begin{eqnarray}
A_\mu(x,\theta)&=&A_\mu^{(0)}(x)+A_\mu^{(1)}(x)+A_\mu^{(2)}(x)+\cdots,\nonumber\\[2mm]
{A'}_\mu(x,\theta)&=&-A_\mu(x,-\theta)=-A_\mu^{(0)}(x)+A_\mu^{(1)}(x)-A_\mu^{(2)}(x)+\cdots,\nonumber\\[2mm]
F_{\mu\nu}(x,\theta)&=&F^{(0)}_{\mu\nu}(x)+F^{(1)}_{\mu\nu}(x)+F^{(2)}_{\mu\nu}(x)+\cdots,\nonumber\\[2mm]
{F'}_{\mu\nu}(x,\theta)&=&-F^{(0)}_{\mu\nu}(x)+F^{(1)}_{\mu\nu}(x)-F^{(2)}_{\mu\nu}(x)+-\cdots,\nonumber\\[2mm]
\psi(x,\theta)&=&\psi^{(0)}(x)+\psi^{(1)}(x)+\psi^{(2)}(x)+\cdots,
\end{eqnarray}
where $\psi$ stands for all 8 spinors under consideration.
The reason why we are led to the $\theta$-expansion
even if we based our NCWS for leptons in Ref.11)
on the Lorentz non-covariant algebra
$[{\hat x}^\mu,{\hat x}^\nu]=i\theta^{\mu\nu}$, leading to fields
not explicitly containing $\theta$,
comes from the demand that
the variational principle should be applied in various cases exhibited in (34).
We shall not repeat the argument but only note that
there is a formidable increase in unknown local fields
in the above expansion.
Namely, although the first terms in the expansion are identified with
the local fields
in the commutative limit,
$A_\mu(x)=A_\mu^{(0)}(x), \psi(x)=\psi^{(0)}(x),
F_{\mu\nu}(x)=\partial_\mu A_\nu(x)-\partial_\nu A_\mu(x)
=F_{\mu\nu}^{(0)}(x)$, 
there occur many unknown local fields in higher order terms.
Nonetheless, gauge invariance of the action is proved$^{[11]}$
order by order up to $n=2$.
\\
\indent
On the other hand, the authors in Re.14)
developed a bottom-up version of NCGT.
They also implicitly assumed $\theta$ dependence of
the field in Lorentz non-invariant NCGT
and introduced additional parameter $h$ to
justify their formulation of NCGT using $h$-expansion
(equivalent to $\theta$-expansion) and
Seiberg-Witten map$^{[15]}$.
They showed that NCGT though violates Lorentz invariance
can be formulated for arbitrary gauge group
including $SU(n)$ by employing Seiberg-Witten map$^{[15]}$ 
to determine all higher-order terms in terms of only the lowest-order term
with due consideration on the gauge parameter coming from the consistency
condition.
Thus they expand the gauge parameter, too,
\begin{eqnarray}
\alpha(x,\theta)&=&\alpha^{(0)}(x)+\alpha^{(1)}(x)+\alpha^{(2)}(x)+\cdots,\nonumber\\[2mm]
\alpha(x,-\theta)&=&\alpha^{(0)}(x)-\alpha^{(1)}(x)+\alpha^{(2)}(x)-+\cdots,
\end{eqnarray}
where $\alpha(x)=\alpha^{(0)}(x)$ is the same as in (29).
The $\theta$-expansion of fields must be compatible with
the $\theta$-expansion of the action
\begin{equation}
{\hat S}_{QED}=
S_{QED}^{(0)}+S_{QED}^{(1)}+S_{QED}^{(2)}+\cdots,
\end{equation}
in the following respect.
The action ${\hat S}_{QED}$ is a functional of
$A_\mu(x,\theta), {A'}_\mu(x,\theta)$ and the spinors
$\chi_1(x,\theta),\cdots,\psi_6(x,\theta)$.
It is invariant under the
NC gauge transformations
(25), (30) and (31).
We now assume following Ref. 14) that
each term in the above expansion
is a local functional of only 4-dimensional fields,
$A_\mu(x), \chi_1(x),\cdots,\psi_6(x)$
with respected gauge invariance, (29) and
\begin{equation}
A_\mu(x)\to ^g\!\!A_\mu(x)=\frac 2e\partial_\mu \alpha(x).
\end{equation}
Hence the $\theta$-expansion (61)
contains both kinds of gauge invariance.
This requires$^{[14]}$ that
there exists the Seiberg-Witten map$^{[15]}$
\begin{eqnarray}
A_\mu(x,\theta)&=&A_\mu[A(x),\theta],\nonumber\\[2mm]
{A'}_\mu(x,\theta)&=&{A'}_\mu[A(x),\theta],\nonumber\\[2mm]
\psi(x,\theta)&=&\psi[A(x),\psi(x),\theta],\nonumber\\[2mm]
U(x,\theta)&=&U[A(x),U(x)\equiv e^{i\alpha(x)},\theta],
\end{eqnarray}
where $\psi$ stands for any spinor $\chi_1(x,\theta),\cdots,\psi_6(x,\theta)$,
such that
\begin{eqnarray}
^{\hat g}\!A_\mu(x,\theta)&=&A_\mu[^g\!A(x),\theta],\nonumber\\[2mm]
^{\hat g}\!{A'}_\mu(x,\theta)&=&{A'}_\mu[^g\!A(x),\theta],\nonumber\\[2mm]
^{\hat g}\psi(x,\theta)&=&\psi[^g\!A(x),^g\!\psi(x),\theta],
\end{eqnarray}
hold true.
This mapping must satisfy the consistency condition
\begin{eqnarray}
^{({\hat g_1}{\hat g_2})}\!A_\mu(x,\theta)&=&^{\hat g_1}(^{\hat g_2}\!A_\mu(x,\theta)),\nonumber\\[2mm]
^{({\hat g_1}{\hat g_2})}\!{A'}_\mu(x,\theta)&=&^{\hat g_1}(^{\hat g_2}\!{A'}_\mu(x,\theta)),\nonumber\\[2mm]
^{({\hat g_1}{\hat g_2})}\!\psi(x,\theta)&=&^{\hat g_1}(^{\hat g_2}\!\psi(x,\theta)).
\end{eqnarray}
\indent
In the following we shall determine
the Seiberg-Witten map
only in infinitesimal gauge transformation,
\begin{eqnarray}
U(x,\theta)&=&(e^{i\alpha(x,\theta)})_*
=1+i\alpha(x,\theta),\nonumber\\[2mm]
\delta_{{\hat\alpha}}A_\mu(x,\theta)&=&
i(\alpha(x,\theta)*A_\mu(x,\theta)-
A_\mu(x,\theta)*\alpha(x,\theta))
+\frac 2e\partial_\mu\alpha(x,\theta),\nonumber\\[2mm]
\delta_{{\hat\alpha}}{A'}_\mu(x,\theta)&=&
-i(\alpha(x,-\theta)*{A'}_\mu(x,\theta)-
{A'}_\mu(x,\theta)*\alpha(x,-\theta))
-\frac 2e\partial_\mu\alpha(x,-\theta),\nonumber\\[2mm]
\delta_{{\hat\alpha}}\chi_1(x,\theta)&=&
i\alpha(x,\theta)*\chi_1(x,\theta)-\chi_1(x,\theta)*i\alpha(x,\theta),\nonumber\\[2mm]
\delta_{{\hat\alpha}}\psi_1(x,\theta)&=&
i\alpha(x,\theta)*\psi_1(x,\theta),\nonumber\\[2mm]
\delta_{{\hat\alpha}}\psi_6(x,\theta)&=&
-i\alpha(x,-\theta)*\psi_6(x,\theta)-\psi_6(x,\theta)*i\alpha(x,\theta).
\end{eqnarray}
The spinors we shall consider below are only $\chi_1$, $\psi_1$ and
$\psi_6$. The other cases will be obvious.
The infinitesimal form of (64) read
\begin{eqnarray}
\delta_{{\hat\alpha}}A_\mu(x,\theta)&=&\delta_\alpha A_\mu(x,\theta)
\equiv A_\mu[A(x)+\delta_\alpha A(x),\theta]-A_\mu[A(x),\theta],\nonumber\\[2mm]
\delta_{{\hat\alpha}}{A'}_\mu(x,\theta)&=&\delta_\alpha {A'}_\mu(x,\theta)
\equiv {A'}_\mu[A(x)+\delta_\alpha A(x),\theta]-{A'}_\mu[A(x),\theta],\nonumber\\[2mm]
\delta_{{\hat\alpha}}\psi(x,\theta)&=&\delta_\alpha\psi(x,\theta)
\equiv \psi[A(x)+\delta_\alpha A(x),\psi(x)+\delta_\alpha \psi(x),\theta]-\psi[A(x),\psi(x),\theta],
\end{eqnarray}
while the consistency condition (65) reads
\begin{equation}
i\delta_\alpha\beta[A,\theta]-
i\delta_\beta\alpha[A,\theta]-
\beta[A,\theta]*\alpha[A,\theta]+\alpha[A,\theta]*\beta[A,\theta]=0.
\end{equation}
Here we define, writing $\alpha(x,\theta)=\alpha[A(x),\theta]=\alpha[A,\theta]$,
\begin{equation}
\delta_\alpha\beta[A,\theta]\equiv\beta[A+\delta_\alpha A,\theta]
-\beta[A,\theta].
\end{equation}
\indent
Our purpose of this section is to show that
the $\theta$ expansion (59) of fields $A_\mu(x,\theta)$ 
and ${A'}_\mu(x,\theta)$  has a consistent solution given
the gauge transformations (66) thanks to the
expansion (60).
We now know$^{[14]}$ that the solution to the consistency condition 
(68) determines $\alpha^{(n)}(x)$ iteratively.
The result up to $n=2$ is as follows. (In the following, we omit
argument $(x)$ for simplicity.)
\begin{eqnarray}
\alpha^{(0)}&=&\alpha,\nonumber\\[2mm]
\alpha^{(1)}&=&\frac e4\theta^{\rho\sigma}\partial_\rho\alpha A_\sigma,\nonumber\\[2mm]
\alpha^{(2)}&=&-\frac {e^2}8\theta^{\rho\sigma}\theta^{\lambda\tau}
\partial_\rho\alpha A_\lambda\partial_\tau A_\sigma.
\end{eqnarray}
It can be shown that
the same 
solution on the gauge fields $A_\mu^{(1)}$ and
$A_\mu^{(2)}$ comes from {\it both}
$\delta_{{\hat\alpha}}A_\mu(x,\theta)$ {\it and} $\delta_{{\hat\alpha}}{A'}_\mu(x,\theta)$
in (66) due to (60). Thus we have
\begin{eqnarray}
A_\mu^{(0)}&=&A_\mu,\nonumber\\[2mm]
A_\mu^{(1)}&=&-\frac e4\theta^{\rho\sigma}A_\rho(\partial_\sigma A_\mu+
F_{\sigma\mu}),\nonumber\\[2mm]
A_\mu^{(2)}&=&\frac {e^2}8\theta^{\rho\sigma}\theta^{\lambda\tau}
(A_\rho A_\lambda\partial_\tau F_{\sigma\mu}
-\partial_\sigma A_\mu\partial_\lambda A_\rho A_\tau+A_\rho F_{\sigma\lambda}F_{\tau\mu}).
\end{eqnarray}
As for the field strength, we find
\begin{eqnarray}
F^{(0)}_{\mu\nu}&=&F_{\mu\nu},\nonumber\\[2mm]
F^{(1)}_{\mu\nu}&=&\frac e2\theta^{\rho\sigma}F_{\mu\rho}F_{\nu\sigma}
-\frac e2\theta^{\rho\sigma}A_\rho\partial_\sigma F_{\mu\nu},\nonumber\\[2mm]
F^{(2)}_{\mu\nu}&=&\frac {e^2}8\theta^{\rho\sigma}\theta^{\lambda\tau}
f_{\mu\nu\rho\sigma\lambda\tau},\nonumber\\[2mm]
f_{\mu\nu\rho\sigma\lambda\tau}&=&F_{\rho\nu}F_{\sigma\lambda}F_{\tau\mu}
-F_{\rho\mu}F_{\sigma\lambda}F_{\tau\nu}\nonumber\\[2mm]
&&+A_\rho(2F_{\mu\lambda}\partial_\sigma F_{\tau\nu}
-2F_{\nu\lambda}\partial_\sigma F_{\tau\mu}
+F_{\sigma\lambda}\partial_\tau F_{\mu\nu}-
\partial_\lambda F_{\mu\nu}\partial_\sigma A_\tau)\nonumber\\[2mm]
&&+A_\rho A_\lambda\partial_\tau\partial_\sigma F_{\mu\nu}.
\end{eqnarray}
Consequently, NC Maxwell action (40) 
has the expansion
\begin{eqnarray}
S^{(0)}_M&=&-\frac 14\int\!d^4xF_{\mu\nu}F^{\mu\nu},\nonumber\\[2mm]
S^{(1)}_M&=&0,\nonumber\\[2mm]
S^{(2)}_M&=&-\frac {e^2}{16}\int\!d^6\theta W(\theta)\theta^{\rho\sigma}\theta^{\lambda\tau}\int\!d^4x
{\tilde f}_{\rho\sigma\lambda\tau},\nonumber\\[2mm]
&&{\tilde f}_{\rho\sigma\lambda\tau}=F_{\mu\rho}F_{\nu\sigma}F^\mu_{\;\;\lambda}F^\nu_{\;\;\tau}
+F^{\mu\nu}(F_{\rho\nu}F_{\sigma\lambda}F_{\tau\mu}
-F_{\rho\mu}F_{\sigma\lambda}F_{\tau\nu})\nonumber\\[2mm]
&&\quad\;\quad\; -F_{\rho\sigma}F^{\mu\nu}F_{\mu\lambda}F_{\nu\tau}\nonumber\\[2mm]
&&\quad\;\quad\; +\frac 12 F_{\rho\sigma}A_\lambda F_{\mu\nu}\partial_\tau F^{\mu\nu}
+A_\rho F^{\mu\nu}F_{\sigma\lambda}\partial_\tau F_{\mu\nu}.
\end{eqnarray}
Thus $C'$-invariance excludes three-photon vertices.
If we employ NC Maxwell action
\begin{eqnarray}
{\hat S}'_M&=&-\frac 14\int\!d^4xd^6\theta W(\theta)
F_{\mu\nu}(x,\theta)*F^{\mu\nu}(x,\theta),
\end{eqnarray}
we have ${S'}^{(0)}_M=S^{(0)}_M$ and ${S'}^{(2)}_M=S^{(2)}_M$ 
together with
\begin{equation}
{S'}^{(1)}_M=-\frac e4\int\!d^6\theta W(\theta)\theta^{\rho\sigma}
\int\!d^4x[F_{\mu\rho}F_{\nu\sigma}F^{\mu\nu}-\frac 14F_{\rho\sigma}F_{\mu\nu}F^{\mu\nu}].
\end{equation}
It contributes nothing, however, because of (17)
as pointed out by Carlson-Carone-Zobin$^{[7]}$
\footnote
{In the old Lorentz-non-invariant version
without $\theta$ integration
${S''}^{(1)}_M=-\frac e4\theta^{\rho\sigma}
\int\!d^4x[F_{\mu\rho}F_{\nu\sigma}F^{\mu\nu}-\frac 14F_{\rho\sigma}F_{\mu\nu}F^{\mu\nu}]$
makes a nontrivial contribution.
}.
In Ref. 7) (17) was derived from the Lorentz invariance, while
we derived it from DFR algebra itself.
\\
\indent
Finally, we give Seiberg-Witten map for the spinors,
\begin{eqnarray}
\chi_1^{(1)}&=&-\frac e2\theta^{\rho\sigma}A_\rho\partial_\sigma\chi_1^{(0)},\nonumber\\[2mm]
\chi_1^{(2)}&=&\frac 18\theta^{\rho\sigma}\theta^{\lambda\tau}
[-ie\partial_\rho A_\lambda\partial_\sigma\partial_\tau\chi_1^{(0)}
-e^2\partial_\rho A_\lambda A_\tau\partial_\sigma\chi_1^{(0)}\nonumber\\[2mm]
&&\quad\quad
-2e^2A_\lambda F_{\rho\tau}\partial_\sigma\chi_1^{(0)}
+e^2A_\lambda A_\rho\partial_\sigma\partial_\tau\chi_1^{(0)}],\nonumber\\[2mm]
\psi_1^{(1)}&=&-\frac e4\theta^{\rho\sigma}A_\rho\partial_\sigma\psi_1^{(0)},\nonumber\\[2mm]
\psi_1^{(2)}&=&\frac 1{32}\theta^{\rho\sigma}\theta^{\lambda\tau}
[-2ie\partial_\rho A_\lambda\partial_\sigma\partial_\tau\psi_1^{(0)}
+e^2A_\rho A_\lambda\partial_\sigma\partial_\tau\psi_1^{(0)}\nonumber\\[2mm]
&&\quad\quad +e^2A_\rho F_{\sigma\lambda}\partial_\tau\psi_1^{(0)}
+2e^2A_\rho\partial_\sigma A_\lambda A_\tau\psi_1^{(0)}
-\frac {e^2}2 \partial_\rho A_\lambda\partial_\sigma A_\tau\psi_1^{(0)}\nonumber\\[2mm]
&&\quad\quad +ie^3 A_\rho A_\tau \partial_\lambda A_\sigma\psi_1^{(0)}
-i\frac {e^3}2 A_\rho A_\tau \partial_\sigma A_\lambda\psi_1^{(0)}],\nonumber\\[2mm]
\psi_6^{(1)}&=&0,\nonumber\\[2mm]
\psi_6^{(2)}&=&\frac 1{32}\theta^{\rho\sigma}\theta^{\lambda\tau}
[4ie\partial_\rho A_\lambda\partial_\sigma\partial_\tau\psi_6^{(0)}
-2e^2\partial_\rho A_\lambda\partial_\sigma\partial_\tau\psi_6^{(0)}\nonumber\\[2mm]
&&\quad\quad\; +4e^2\partial_\sigma A_\lambda A_\rho \partial_\tau\psi_6^{(0)}].
\end{eqnarray}
Dirac action for the spinor $\chi_1$ has the expansion with upper index still attached
\begin{eqnarray}
{\hat S}_D&=&
S_D^{(0)}+S_D^{(1)}+S_D^{(2)}+\cdots,\nonumber\\[2mm]
S_D^{(0)}&=&\int\!d^4x{\bar\chi}_1^{(0)}i\gamma^\mu\partial_\mu\chi_1^{(0)},\nonumber\\[2mm]
S_D^{(1)}&=&
\frac{ie}2\int\!d^6\theta W(\theta)\theta^{\rho\sigma}\int\!d^4x{\bar\chi}^{(0)}_1
i\gamma^\mu F_{\rho\mu}\partial_\sigma\chi^{(0)}_1,\nonumber\\[2mm]
S_D^{(2)}&=&\int\!d^4xd^6\theta W(\theta)
[{\bar\chi}^{(1)}_1i\gamma^\mu \partial_\mu\chi^{(1)}_1
+{\bar\chi}^{(0)}_1i\gamma^\mu \partial_\mu\chi^{(2)}_1
+{\bar\chi}^{(2)}_1i\gamma^\mu \partial_\mu\chi^{(0)}_1\nonumber\\[2mm]
&&+{\bar\chi}^{(0)}_1\frac {ie}2\theta^{\rho\sigma}(i\gamma^\mu\partial_\rho A_\mu)
\partial_\sigma\chi^{(1)}_1
+{\bar\chi}^{(1)}_1\frac {ie}2\theta^{\rho\sigma}(i\gamma^\mu\partial_\rho A_\mu)
\partial_\sigma\chi^{(0)}_1\nonumber\\[2mm]
&&+{\bar\chi}^{(0)}_1\frac {ie}2\theta^{\rho\sigma}
(i\gamma^\mu\partial_\rho A_\mu^{(1)})\partial_\sigma\chi^{(0)}_1].
\end{eqnarray}
The second term $S_D^{(1)}$ vanishes due to (17).
Consequently we do not need evaluate $S_D^{(1)}$ in the following.
For $\psi_1$, we get
\begin{eqnarray}
{\hat S}_D&=&
S_D^{(0)}+S_D^{(2)}+\cdots,\nonumber\\[2mm]
S_D^{(0)}&=&
\int\!d^4x{\bar\psi}^{(0)}_1i\gamma^\mu D_\mu\psi^{(0)}_1,\;\;\;
D_\mu=\partial_\mu-\frac {ie}2A_\mu, \nonumber\\[2mm]
S_D^{(2)}&=&\int\!d^4xd^6\theta W(\theta)\{{\bar\psi}^{(1)}_1i\gamma^\mu D_\mu\psi^{(1)}_1
+{\bar\psi}^{(0)}_1i\gamma^\mu D_\mu\psi^{(2)}_1
+{\bar\psi}^{(2)}_1i\gamma^\mu D_\mu\psi^{(0)}_1\nonumber\\[2mm]
&&+\frac e2[-\frac 18\theta^{\rho\sigma}\theta^{\lambda\tau}
{\bar\psi}^{(0)}_1(i\gamma^\mu\partial_\rho\partial_\lambda A_\mu)
\partial_\sigma\partial_\tau\psi^{(0)}_1\nonumber\\[2mm]
&&\quad\; +{\bar\psi}^{(0)}_1\frac i2\theta^{\rho\sigma}(i\gamma^\mu\partial_\rho A_\mu)
\partial_\sigma\psi^{(1)}_1
+{\bar\psi}^{(1)}_1\frac i2\theta^{\rho\sigma}(i\gamma^\mu\partial_\rho A_\mu)
\partial_\sigma\psi^{(0)}_1\nonumber\\[2mm]
&&\quad\; +{\bar\psi}^{(0)}_1\frac i2\theta^{\rho\sigma}
(i\gamma^\mu\partial_\rho A_\mu^{(1)})\partial_\sigma\psi^{(0)}_1\nonumber\\[2mm]
&&\quad\; +{\bar\psi}^{(1)}_1i\gamma^\mu A_\mu^{(1)}\psi^{(0)}_1
+{\bar\psi}^{(0)}_1i\gamma^\mu A_\mu^{(1)}\psi^{(1)}_1
+{\bar\psi}^{(0)}_1i\gamma^\mu A_\mu^{(2)}\psi^{(0)}_1]\}.
\end{eqnarray}
For $\psi_6$ we find
\begin{eqnarray}
{\hat S}_D&=&
S_D^{(0)}+S_D^{(2)}+\cdots,\nonumber\\[2mm]
S_D^{(0)}&=&
\int\!d^4x{\bar\psi}^{(0)}_6i\gamma^\mu D_\mu\psi^{(0)}_6,\;\;\;
D_\mu=\partial_\mu+ieA_\mu, \nonumber\\[2mm]
S_D^{(2)}&=&\int\!d^4x\int\!d^6\theta W(\theta)[{\bar\psi}^{(1)}_6i\gamma^\mu D_\mu\psi^{(1)}_6
+{\bar\psi}^{(0)}_6i\gamma^\mu D_\mu\psi^{(2)}_6
+{\bar\psi}^{(2)}_6i\gamma^\mu D_\mu\psi^{(0)}_6\nonumber\\[2mm]
&&\quad\; +\frac e8\theta^{\rho\sigma}\theta^{\lambda\tau}
{\bar\psi}^{(0)}_6(i\gamma^\mu\partial_\rho\partial_\lambda A_\mu)
\partial_\sigma\partial_\tau\psi^{(0)}_6\nonumber\\[2mm]
&&\quad\; +\frac {ie}2{\bar\psi}^{(0)}_6\theta^{\rho\sigma}(i\gamma^\mu\partial_\rho A_\mu^{(1)})
\partial_\sigma\psi^{(0)}_6
-e{\bar\psi}^{(0)}_6i\gamma^\mu A_\mu^{(2)}\psi^{(0)}_6].
\end{eqnarray}
To put ${\hat S}_D^{(2)}$ into a final form
(76) should be used.
We shall not try in this paper to do phenomenological calculations based on the above Seiberg-Witten map.
It would be enough to comment that
Carlson-Carone-Zobin$^{[7]}$ calculated $2\gamma\to 2\gamma$ scattering
based on (73)
and found a distinctive deviation from the
standard model result.
\section{Conclusions}
\indent
As the first step toward formulating NCSM
by considering the total fermion field as a NC bi-module,
we have constructed in this paper Lorentz-invariant NCQED.
All possible spinors are considered,
which are reduced to one neutral and four charged spinors
in the commutative limit. This charge quantization
is tight enough that
it precisely gives the correct hypercharge assignment
of leptons in addition to their electric charge. 
\\
\indent
The important aspect of our NCQED is its Lorentz invariance.
It was first formulated by Carlson, Carone and Zobin$^{[7]}$.
The commutative limit of the Lorentz-invariant NCQED
{\it smoothly} coincides with Lorentz-invariant QED.
\\
\indent
The oscillating damping factor for nonplanar diagrams
first observed in Ref. 17) is replaced with an invariant damping factor.
Moreover, the singular behavior of Green functions at $\theta\to 0$
found in the literature$^{[10],[13]}$
may be evaded using a new UV limit, assumed to be valid
near IR region indistinguishable from the commutative limit,
in an {\it invariant} way.
This conjecture was confirmed 
in the proper self-energy diagram
of NC scalar model in the one-loop approximation.
\\
\indent
We define covariant operator fields on DFR algebra (8)
and associate them
with $c$-number Weyl symbols which enjoy the same Lorentz
covariance. The latter are field quantities to be subsequently quantized.
Even in the commutative limit, DFR algebra remains intact.
Our Minkowski space-time is a parameter space like
the 4-dimensional phase space.
If we say that
our Minkowski space-time becomes non-commutative at very short distances
by assuming the commutator $[{\hat x}^\mu,{\hat x}^\nu]=i\theta^{\mu\nu}$
for constant $\theta$,
we immediately sacrifice the Lorentz symmetry,
which could never be remedied as far as we stick to constant $\theta$ algebra.
If we employ Lorentz-covariant algebra
(8) following Carlson, Carone and Zobin$^{[7]}$,
the deformation parameter becomes an integration variable. 
Nonetheless, it is possible to study small $\theta$ 
using Seiberg-Witten map because it is dimensionfull.
In addition, one-loop calculation in the section 4
indicates how to obtain invariant amplitudes on non-commutative space-time.
\\
\indent
Finally, we would like to point out that,
since QED is not a closed theory but
is unified with weak interactions at present energy,
NCQED should also be a part of a larger theory.
This subject will be a theme in the following papers.
\section*{Acknowledgements}
The author is grateful to
H. Kase, Y. Okumura and E. Umezawa for
useful discussions and continuous encouragement.
\appendix
\section{Derivative operator for constant $\theta$}
The derivative operator for the field
${\hat \varphi}({\hat x})$ on the non-commutative space-time
$[{\hat x}^\mu,{\hat x}^\nu]=i\theta^{\mu\nu}$
is defined by
$$
{\check p}_\mu=-i\theta_{\mu\nu}{\hat x}^\nu,
$$
where $\theta_{\mu\nu}$ is the inverse of the matrix $\theta^{\mu\nu}$,
$\theta_{\mu\nu}\theta^{\nu\lambda}=\delta_\mu^{\;\,\lambda}$
so that $[{\check p}_\mu,{\hat x}^\nu]=\delta_\mu^{\;\,\nu}$.
It can be shown that the commutator
$[{\check p}_\mu,{\hat \varphi}({\hat x})]$ equals (13) with
${\hat \theta}$ disregarded.
There are two problems, here. First of all
$\theta^{\mu\nu}$ must be assumed to be invertible.
The second is that the commutative limit
is not smooth as pointed out in the Appendix A in Ref.11).
Suppose that the matrix $(\theta^{\mu\nu})$ is invertible
and put into the canonical form with only non vanishing elements $\theta_{1,2}$
such that ${\hat x}^1$ and ${\hat x}^3$ are diagonalized with the basis $|x^1,x^3\rangle$.
Then ${\check p}_0=(i/\theta_1){\hat x}^1$ and ${\check p}_2=(i/\theta_2){\hat x}^3$
become singular in the commutative limit
$\theta_1, \theta_2\to 0$,
although the commutator $[{\check p}_\mu,{\hat \varphi}({\hat x})]$ for any $\mu$
is well-defined.
On the other hand, if all coordinates commute,
they can be simultaneously diagonalized with 
the different basis $|x^\mu\rangle$.
In this case, we simply put
${\check p}_\mu=\partial/\partial x^\mu$.
This non-smoothness no longer bothers us
for DFR algebra.
\section{Operator product and Moyal product}
In this section we shall give the proof of (20).
The integrand ${\tilde \varphi}_{12}(k,\sigma)$
is calculated, using the notation
$(k\times k'){\hat \theta}=(k\times k')_{\mu\nu}{\hat \theta}^{\mu\nu}$
with $(k\times k')_{\mu\nu}\equiv(1/2)(k_{1\mu}k'_{2\nu}-k_{1\nu}k'_{2\mu})$
and ${\hat T}(k,\sigma)=e^{ik{\hat x}+i\sigma{\hat \theta}}$,
as follows.
\begin{eqnarray*}
{\hat \varphi}_1({\hat x},{\hat \theta}){\hat \varphi}_2({\hat x},{\hat \theta})
&=&\frac 1{(2\pi)^8}\int\!d^4kd^6\sigma{\tilde \varphi}_1(k,\sigma){\hat T}(k,\sigma)
\int\!d^4k'd^6\sigma'{\tilde \varphi}_2(k',\sigma'){\hat T}(k',\sigma')\nonumber\\[2mm]
&=&\frac 1{(2\pi)^8}\int\!d^4kd^6\sigma d^4k'd^6\sigma'{\tilde \varphi}_1(k,\sigma){\tilde \varphi}_2(k',\sigma')
e^{-{\mbox{\tiny$\frac{i}{2}$}(k\times k'){\hat\theta}}}
{\hat T}(k+k',\sigma+\sigma')\nonumber\\[2mm]
&=&\frac 1{(2\pi)^8}\int\!d^4Kd^6\Sigma'd^4k'd^6\sigma'
{\tilde \varphi}_1(K-k',\Sigma'-\sigma')\nonumber\\[2mm]
&&\times{\tilde \varphi}_2(k',\sigma'){\hat T}(K,\Sigma'-\frac 12K\times k')
\nonumber\\[2mm]
&=&\frac 1{(2\pi)^8}\int\!d^4Kd^6\Sigma {\tilde \varphi}_{12}(K,\Sigma){\hat T}(K,\Sigma),\nonumber\\[2mm]
{\tilde \varphi}_{12}(K,\Sigma)&=&\frac 1{(2\pi)^4}\int\!d^4k'd^6\sigma'
{\tilde \varphi}_1(K-k',\Sigma+\frac 12 K\times k'-\sigma')
{\tilde \varphi}_2(k',\sigma').
\end{eqnarray*}
Hence, we have
\begin{eqnarray*}
\varphi_{12}(x,\theta)&=&
\frac 1{(2\pi)^4}\int\!d^4Kd^6\Sigma e^{iKx+i\Sigma\theta}{\tilde \varphi}_{12}(K,\Sigma)\nonumber\\[2mm]
&=&\frac 1{(2\pi)^{20}}\int\!d^4Kd^6\Sigma
d^4k'd^6\sigma'e^{iKx+i\Sigma\theta}
\int\!d^4x_1d^6\theta_1d^4x_2d^6\theta_2\varphi_1(x_1,\theta_1)
\varphi_2(x_2,\theta_2)\nonumber\\[2mm]
&&\times e^{-i(K-k')x_1-i(\Sigma+\frac 12K\times k'-\sigma')\theta_1}
    e^{-ik'x_2-i\sigma'\theta_2}\nonumber\\[2mm]
&=&\frac 1{(2\pi)^{20}}\int\!d^4Kd^6\Sigma d^4k'd^6\sigma' 
d^4x_1d^6\theta_1d^4x_2d^6\theta_2 
\nonumber\\[2mm]
&&\times e^{iKx+i\Sigma\theta-i(K-k')x_1-i(\Sigma-\sigma')\theta_1
-ik'x_2-i\sigma'\theta_2}
(e^{\frac i2(\partial_1\times\partial_2)\theta_1}
\varphi_1(x_1,\theta_1)\varphi_2(x_2,\theta_2))\nonumber\\[2mm]
&=&e^{\frac i2\theta^{\mu\nu}\frac{\partial}{\partial x^\mu}
\frac{\partial}{\partial y^\nu}}
\varphi_1(x,\theta)\varphi_2(y,\theta)|_{x=y}\nonumber\\[2mm]
&\equiv&\varphi_1(x,\theta)*\varphi_2(x,\theta).
\end{eqnarray*}
This is nothing but (20).
\vspace{3mm}


\begin{thebibliography}{99}
\vspace{3mm}
\bibitem{1}
M. Chaichian, P. Pre${\check {\rm s}}$najder, M. M. Sheikh-Jabbari and
A. Tureaunu, `Noncommutative Standard Model:Model Building',
hep-th/0107055(2001).
\bibitem{2}
X. Calmet, B. Jur${\check{\rm c}}$o, P. Schupp, J. Wess and
M. Wohlgenannt,
`The Standard Model on Non-Commutative Space-Time',
hep-th/0111115(2001).
\bibitem{3}
Xiao-Gang He,
`Strong, Electroweak Interactions and Their Unification with Non-Commutative Space-Time',
hep-th/0202223(2002).\\
See also, I. Hinchliffe and N. Kersting, `Review of the Phenomenology of Noncommutative
Geometry', hep-th/0205040(2002).
\bibitem{4}
A.~Connes and J.~Lott, Nucl. Phys. Proc. Suppl. {\bf 18B}$\,$(1990), 29.
\bibitem{5}
A.~Connes, {\it Noncommutative Geometry}$\;$
       (Academic Press, New York, 1994), Chapter 6.\\
A. Connes, J. Math. Phys. {\bf 36}$\,$(1995), 6194;
Commun. Math. Phys. {\bf 1182}$\,$(1996), 155.
\bibitem{6}
H. Kase, M. Morita and Y. Okumura, Intern. Jour. Modern Physics,
{\bf A 16}$\,$(2001), 3203.
\bibitem{7}
C. E. Carlson, C. D. Carone and N. Zobin,
`Noncommutative Gauge Theory without Lorentz Violation',
hep-th/0206035(2002).
\bibitem{8}
H. Kase, M. Morita and Y. Okumura, Prog. Theor. Phys. {\bf 102}$\,$(1999), 1027.\\
Connes' $U(1)$ may be called super-compact,
to be compared with compact $U(1)$ associated with the observed charge
quantization, C. N. Yang, Phys. Rev. {\bf D8}$\,$(1970), 2360.
\bibitem{9}
I. S. Sogami, Prog. Theor. Phys. {\bf 94}$\,$(1995), 117.
\bibitem{10}
M. Hayakawa, `Perturbative Analysis on Infra-red and Ultraviolet Aspects
of Noncommutative QED on ${\bf R}^4$', hep-th/9912167(1999); Phys. Lett. {\bf B478}(2000), 394.
\bibitem{11}
H. Kase, M. Morita and Y. Okumura, Prog. Theor. Phys. {\bf 106}$\,$(2001), 187.
\bibitem{12}
S. Doplicher, K. Fredenhagen and J. E. Roberts, Phys. Lett. {\bf B331}$\,$(1994),
39;$\,$Commun. Math. Phys. {\bf 172}$\,$(1995), 187.
\bibitem{13}
See, for instance, S. Minwalla, M. Van Raamstonk and N. Seiberg,
`Noncommutative perturbative dynamics', JHEP, {\bf 0002}(2000), 020;
hep-th/9912072(1999).
\bibitem{14}
B. Jur${\check{\rm c}}$o, L. M\"oller, S. Shraml, P. Schupp, and J. Wess,
`Construction of non-Abelian gauge theories on noncommutative spaces',
hep-th/0104153(2001).
\bibitem{15}
N. Seiberg and E. Witten,  `String Theory and Noncommutative Geometry'
J. High Energy Phys.  {\bf 9909}(1999), 032, hep-th/9908142(1999).
\bibitem{16}
H. S. Snyder, Phys. Rev. {\bf 71}(1947), 38.
\bibitem{17}
T. Filk, Phys. Lett. {\bf B376} (1996), 53.
\end{thebibliography}
\end{document}